\documentclass[aps,pre,twocolumn,superscriptaddress,longbibliography,nobibnotes,nodoi,nourl,noeprint]{revtex4-2}
\usepackage{amsmath,amssymb,physics,bm,siunitx}
\usepackage{xcolor,graphicx,soul}
\usepackage[colorlinks=true,citecolor=blue,urlcolor=blue,linkcolor=black]{hyperref}
\usepackage{subfiles}

 \def\dd{{\rm d}}

\newcommand{\rhoc}{\rho_{\rm c}}
\newcommand{\nb}{n_{\rm b}}
\newcommand{\mb}{\lceil\nb\sigma\rceil}
\newcommand{\tausol}{\tau_{\rm sol}}
\newcommand{\Nsol}{N_{\rm sol}}
\newcommand{\vsol}{v_{\rm sol}}

\begin{document}
	
\title{Nonequilibrium phase transition in single-file transport at high crowding}

\author{Annika Vonhusen} \email{avonhusen@uos.de}
\affiliation{Universit\"{a}t Osnabr\"{u}ck, Fachbereich
  Mathematik/Informatik/Physik, Barbarastra{\ss}e 7, D-49076
  Osnabr\"uck, Germany}

\author{Sören Schweers} \email{sschweers@uos.de}
\affiliation{Universit\"{a}t Osnabr\"{u}ck, Fachbereich
  Mathematik/Informatik/Physik, Barbarastra{\ss}e 7, D-49076
  Osnabr\"uck, Germany}

\author{Artem Ryabov} \email{rjabov.a@gmail.com} \affiliation{Charles
  University, Faculty of Mathematics and Physics, Department of
  Macromolecular Physics, V Hole\v{s}ovi\v{c}k\'{a}ch 2, CZ-18000
  Praha 8, Czech Republic}

\author{Philipp Maass} \email{maass@uos.de}
\affiliation{Universit\"{a}t Osnabr\"{u}ck, Fachbereich
  Mathematik/Informatik/Physik, Barbarastra{\ss}e 7, D-49076
  Osnabr\"uck, Germany}

\date{November 19, 2025}

\begin{abstract}
Driven particle transport in crowded and confining environments is fundamental to diverse phenomena across physics, chemistry, and biology. A main objective in studying such systems is to identify novel emergent states and phases of collective dynamics.
Here, we report on a nonequilibrium phase transition occurring in periodic structures at high particle densities. This transition separates a weak-current phase of thermally activated transport from a high-current phase of solitary wave propagation. It is reflected also 
in a change of universality classes characterizing correlations of particle current fluctuations.
Our findings demonstrate that sudden changes to high current states can occur when increasing particle densities beyond critical values.
\end{abstract}

\maketitle

\section{Introduction}
Driven particle motion in confined crowded environments is of crucial importance
for many processes in physics, chemistry and biology. Examples are
particle motion in micro- and nanofluidic devices \cite{Wei/etal:2000, Ma/etal:2015, Wang/etal:2024}, optical potentials, 
mesoporous materials \cite{Yiu/etal:2001, Hartmann:2005}, carbon nanotubes \cite{Zeng/etal:2018, Li/Noy:2025} and pores of zeolites \cite{Hahn/etal:1996, VanDeVoorde/Sels:2017, Chmelik/etal:2023, Liu/etal:2025}, ion motion through narrow channels \cite{Hille:2001},
movement of motor proteins and molecular motors \cite{Kolomeisky:2013, Ciandrini/etal:2014, Kolomeisky:2015, Ariga:2024, Grieb/etal:2025}, and
protein synthesis by ribosomes \cite{MacDonald/etal:1968, Erdmann-Pham:2020}.
One-dimensional driven diffusive systems provide a powerful framework for
studying key features of these phenomena.
A main challenge is to model and understand the diverse emergent states of collective dynamics and transitions between them.

Recently, a new solitary cluster-wave mechanism was reported for
transport of microparticles in highly crowded periodic structures
\cite{Antonov/etal:2022a, Cereceda-Lopez/etal:2023}. Here we show that the solitary cluster waves lead to a novel phase transition between a state of thermally activated dynamics and persistent wave propagation. The phase transition is reflected also in correlations between particle current fluctuations.
It occurs in a homogeneous closed system contrary to previously reported nonequilibrium phase transitions in one-dimensional driven diffusive systems, where open boundaries and/or inhomogeneous environments are required
\cite{Krug:1991, Schuetz/Domany:1993, Kolomeisky:1998, Hinsch/Frey:2006, Brzank/Schuetz:2007}.
Exceptions are dynamical phase transitions in large deviations as in Ref.~\cite{Bodineau/Derrida:2005}. 

Specifically, we consider the Brownian asymmetric simple exclusion process (BASEP)  \cite{Lips/etal:2018, Lips/etal:2019}.  In this paradigmatic model of driven single-file diffusion, particles with excluded volume interaction are dragged across a periodic potential. Many intriguing features have been explored for this model and confirmed in experiments: current-density relations strongly change with the particle size, leading to different types of nonequilibrium phase diagrams in open systems coupled to particle reservoirs. Local transition kinetics of a tracer show peculiar faster uphill transitions, and can be used for probing collective effects in the dynamics \cite{Ryabov/etal:2019}. Hydrodynamic interactions lead to an 
effective enhancement of potential barriers in flow-driven system and can induce jamming \cite{Cereceda-Lopez/etal:2021}.

Solitary waves in BASEP occur at high densities due to clustering of particles. 
The solitons manifest themselves as periodic detachment and attachment processes between clusters and allow for
transport of particles over energetic barriers much larger than the thermal energy \cite{Antonov/etal:2025b}. 
Their appearance is restricted to ranges around certain magic particle sizes that widen with the drag force \cite{Antonov/etal:2022a}.
There can be multiple solitons and the number of them is controlled by the filling factor of potential wells.
Measurements have shown an effective repulsive interaction to exist between solitons \cite{Cereceda-Lopez/etal:2023} that
can be explained by a delayed relaxation of particle clusters in the wave propagation process \cite{Antonov/etal:2024}.

\section{Brownian asymmetric simple exclusion process}
In BASEP, hard spheres of diameter $\sigma$ 
are driven by a drag force $f$ across a sinusoidal potential
\begin{equation}
U(x)=\frac{U_0}{2}\cos(\frac{2\pi x}{\lambda})\,.
\end{equation}
The overdamped Brownian motion of $N$ such particles is described by the Langevin equations
\begin{equation}
\dot x_i(t)=\mu\!\left(f-\frac{\dd U}{\dd x_i}\right)+\sqrt{2D}\,\xi_i(t)\,,\hspace{1em}i=1,\ldots,N\,,
\label{eq:langevin}
\end{equation}
subject to the hard-sphere constraints
 \begin{equation}
 |x_i-x_j|\ge\sigma\,.
 \end{equation}
Here, $x_i$ are particle positions, $\xi_i(t)$ are Gaussian processes with zero mean and covariance $\langle\xi_i(t)\xi_j(t')\rangle=\delta_{ij}\delta(t-t')$,
 $\mu$ is the particle mobility, and $D$ the self-diffusion coefficient. 

The system size $L$ is an integer multiple of the wavelength $\lambda$ and 
periodic boundary conditions are considered.
The particle density is
\begin{equation}
\rho=\frac{N}{L}\,,
\end{equation}
and the instantaneous particle current density is
\begin{equation}
\mathcal{J}(t)=\frac{1}{L}\sum_{i=1}^N \dot x_i(t)\,.
\end{equation}
After a transient time, the dynamics become stationary. In this nonequilibrium steady state the current 
is characterized by a mean value $J(\rho)=\langle \mathcal{J}(t)\rangle$.

To solve the  Langevin equations \eqref{eq:langevin}, we
apply the Brownian cluster dynamics method \cite{Antonov/etal:2022c, Antonov/etal:2025a}. This allows for
simulating the dynamics also for $D=0$, i.e.\ in the zero-noise limit. As units of energy, length, and time, 
we take $U_0$, $\lambda$, and $\lambda^2/\mu U_0$, respectively.

\section{Results}
\label{sec:results}
\begin{figure}[t!]
\centering
\includegraphics[width=0.8\columnwidth]{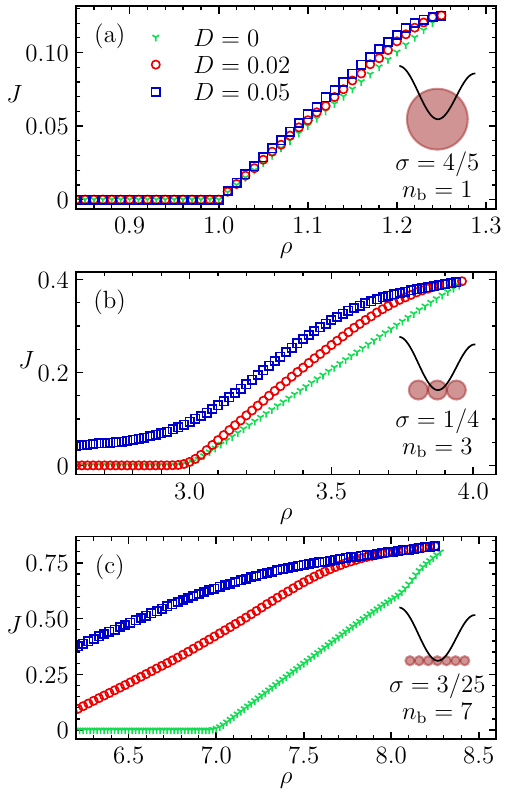}
\caption{\small Particle current $J(\rho)$ as a function of particle density $\rho$ for three particle diameters (a) $\sigma=4/5$
(b) $1/4$, and (c) $3/25$ for drag force $f=0.1$ and three values of noise strength  $D=0$ (zero-noise limit), $0.02$, and $0.05$.
Insets illustrate basic stable clusters of size $\nb$ in a single well of the tilted sinusoidal potential.}
\label{fig:J-rho}
\end{figure}

Figure~\ref{fig:J-rho} shows current-density relations $J(\rho)$ for three particle diameters (a) $\sigma=4/5$, (b) $1/4$, and (c) $3/25$ for 
drag force $f=0.1$ and three noise strengths $D=0$, $0.02$, and $0.05$. It illustrates a phase transition in the zero-noise limit ($D=0$, green symbols) and the effects of thermal noise.
For $\rho$ approaching the maximal particle density  $\rho_{\rm max}=1/\sigma$, the system is completely covered by the particles. They form one big cluster with velocity $\mu f$ then, yielding 
$J_{\rm max}=\rho_{\rm max}\mu f=\mu f/\sigma$ for all particle diameters and noise strengths.

\subsection{Phase transitions in the zero-noise limit}
\label{subsec:zero-noise-limit}
The nonequilibrium phase transition occurs at  a critical particle density $\rhoc$ between a resting and running state: below $\rhoc$ particle transport is absent while above $\rhoc$ the particle current becomes nonzero. 
For $\rho\gtrsim\rhoc$ close to the critical density, the current increases linearly with $\rho$ for all $\sigma$. 

When $\rho\le1$, each potential well can be occupied by one particle and these single particles cannot surmount barriers to neighboring wells in the absence of noise. Accordingly, the current is zero. If $\rho>1$, multiple particles occupy a fraction of wells and clusters of particles in contact form around potential minima. One may expect still no transport to occur because single particles cannot overcome the potential barriers.
Surprisingly, this is not the case. The clusters can mediate particle transport across the potential barriers: 
in a periodic process of
cluster translation, detachment and attachment, 
solitary cluster waves emerge \cite{Antonov/etal:2022a, Antonov/etal:2024}.
Particles of the clusters involved in this process stay together because of the forces from the external
periodic potential $U(x)$. For the cluster to move, the potential for the center of mass of the clusters is decisive and this
can have much smaller or even no barriers.

The critical density $\rhoc$ for the onset of solitary wave propagation is
\begin{equation}
\rhoc=\frac{\nb}{\mb}\,,
\label{eq:rhoc}
\end{equation}
where $\nb$ is the number of particles (size) of a basic cluster, called $n_{\rm b}$-cluster; $\lceil x\rceil$ denotes the smallest integer larger than or equal to $x$. The $n_{\rm b}$-cluster
is the largest mechanically stable one that can form in the potential landscape, and
$\mb$ is the number of potential wells needed for accommodating an $\nb$-cluster. It equals also the spatial period of the soliton motion.
The size $\nb$ follows from a minimal residual free space principle and is related to Euler's totient function of number theory \cite{Antonov/etal:2024, Antonov/etal:2025b}. It depends on the particle diameter $\sigma$ and drag force $f$.

For calculating the current for $\rho>\rhoc$, we resort to recently derived results in Refs.~\cite{Mishra/etal:2025,Antonov/etal:2025b}.
The transport mediated by solitary waves has the remarkable property that the total displacement $\Delta$ of all particles in one soliton period $\tausol$ is one per soliton, i.e.\ one wavelength $\lambda$ of the periodic potential. In the presence of $\Nsol$ solitons, 
the total particle displacement in time $\tausol$ is $\Nsol$. We thus obtain $J=\Delta/L\tausol=\Nsol/L\tausol$. 

The number density  of solitons increases as $\Nsol/L=(\rho-\rhoc)\delta\Nsol$, where $\delta\Nsol$ is a constant increment of the soliton number when one particle is added to the system.
Putting together the unit displacement law,  particle number conservation, and the condition that all potential wells are accommodating clusters,
we have derived $\delta\Nsol=\mb$. Thus the current is
\begin{equation}
J=(\rho-\rhoc)\,\vsol\,,
\label{eq:J-rho}
\end{equation}
where $\vsol=\mb/\tausol$ is the soliton velocity. It gives the slope of the green curve in Fig.~\ref{fig:J-rho} close to $\rhoc$. 

For larger $\rho$,
$\vsol$ depends weakly on $\rho$ due to soliton-soliton interactions \cite{Antonov/etal:2024}.
This dependence is almost negligible for the two larger diameters $\sigma=4/5$ and $1/4$ in Figs.~\ref{fig:J-rho}(a) and \ref{fig:J-rho}(b),
where the linear increase of $J$ with $\rho$ continues until $\rho$ approaches the maximal density $\rho_{\rm max}=1/\sigma$.
For the small diameter $\sigma=3/25$ in Fig.~\ref{fig:J-rho}(c), by contrast,
a kink occurs when the mean distance $L/\Nsol$ between solitons becomes equal to $\mb$, i.e.\ for 
$\rho_\ast=\rhoc+1/\mb^2=8$. After the kink, $J(\rho)$ more steeply increases towards the maximal current $J_{\rm max}$.

In the presence of noise, particles can perform thermally activated transitions over the potential barriers and the current is no longer strictly zero for densities $\rho<\rhoc$.  At high densities, these activated transitions
contribute much less to particle transport than the solitary cluster waves do.
The transition at $\rhoc$ at $D=0$ is clearly reflected also at low noise, as can be seen from the results for $D=0.02$ in Figs.~\ref{fig:J-rho}(a) and (b). It manifests itself in a kink in the current-density relation at $\rhoc$.

Generally, as expected, noise effects become larger for smaller particle diameters. For the largest diameter $\sigma=4/5$
in Fig.~\ref{fig:J-rho}(a), the data at the higher simulated noise strength $D=0.05$ still show the transition, while for the smaller
$\sigma=1/4$ in Fig.~\ref{fig:J-rho}(b) the transition disappears at $D=0.05$, and for $\sigma=3/25$ in Fig.~\ref{fig:J-rho}(c) it is no longer visible even at $D=0.02$.

\begin{figure}[b!]
\centering
\includegraphics[width=0.95\columnwidth]{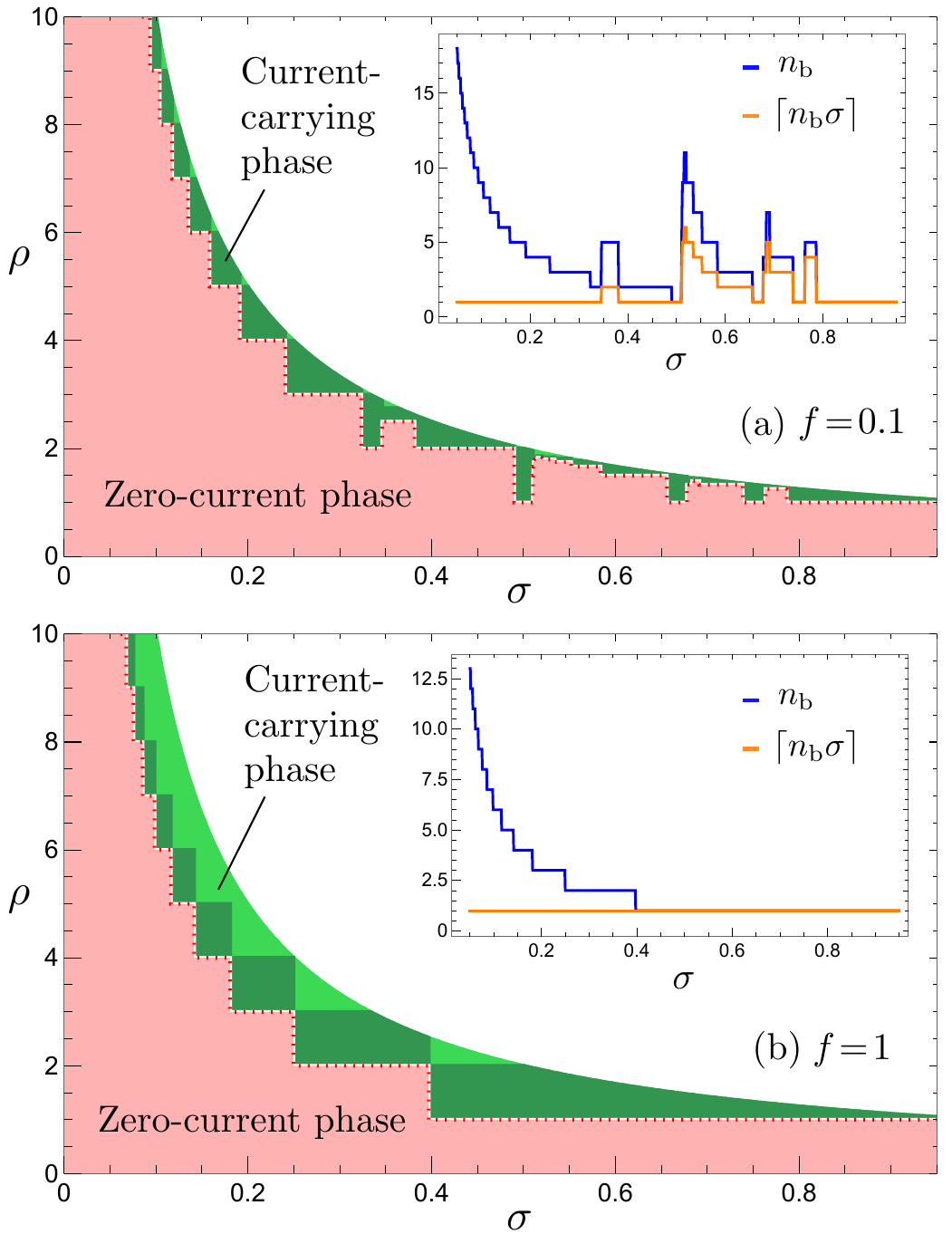}
\caption{\small Phase diagram of jammed (red) and current-carrying phase (green)
for two drag forces (a) $f=0.1$ and (b) $f=1$.
At the transition (dotted line), the particle density equals the critical value $\rhoc=\rhoc(\sigma,f)$, where the dependence on $\sigma$ and $f$ is via $\nb=\nb(\sigma,f)$ in Eq.~\eqref{eq:rhoc}. How $\nb=\nb(\sigma,f)$ and $\mb$ vary with $\sigma$ is shown in the insets for the two $f$ values. 
In the dark green parts of the current-carrying phases, the current increases linearly with $\rho$. At the border between the dark and bright green areas the particle density is $\rho_\ast=\rhoc+1/\mb^2$.
The white region marks a forbidden domain, as $\rho \leq 1/\sigma$ must be satisfied.}
\label{fig:phase-diagrams-fixed-f}
\end{figure}

\begin{figure}[t!]
\centering
\includegraphics[width=0.95\columnwidth]{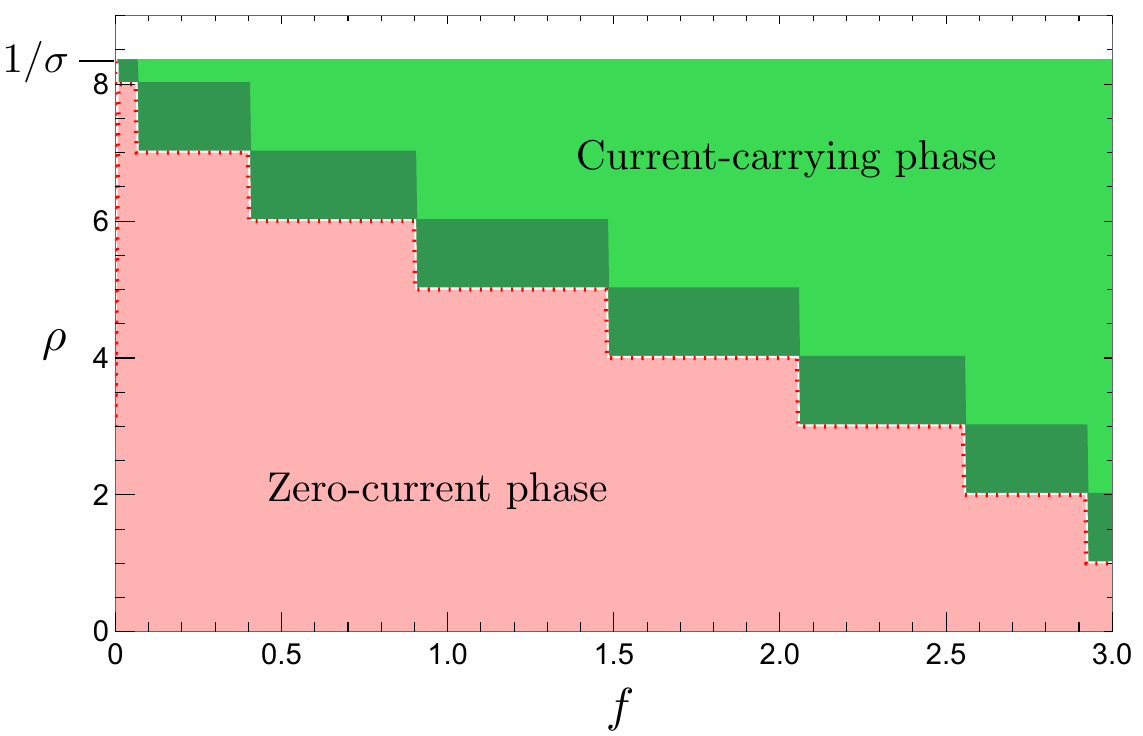}
\caption{\small Phase diagram of jammed (red) and current-carrying phase (green) for fixed particle diameter $\sigma=3/25$. The critical density $\rhoc=\rhoc(\sigma,f)$ at the transition (dotted lines) decreases in a step-wise manner with increasing $f$ due to the decrease of $\nb$ with $f$ (compare also data in the two insets of 
Fig.~\ref{fig:phase-diagrams-fixed-f}). As in Fig.~\ref{fig:phase-diagrams-fixed-f}, the 
current increases linearly with $\rho$ in dark green areas, while beyond the border at $\rho_\ast=\rhoc+1/\mb^2$
the behavior becomes nonlinear.}
\label{fig:phase-diagrams-fixed-sigma}
\end{figure}

\subsection{Phase diagrams}
\label{subsec:phase-diagrams}
Figure~\ref{fig:J-rho} illustrates the transition from a jammed to a current-carrying phase at $\rhoc$  for three particle diameters at drag force $f=0.1$. The critical density $\rhoc$ changes with the particle diameter due the change of the basic cluster size $\nb$ with $\sigma$. 
The full phase diagram in the $\sigma\rho$-plane for $f=0.1$ is displayed in Fig.~\ref{fig:phase-diagrams-fixed-f}(a), where the red area marks the zero-current phase of mechanically stable states and the green areas the current-carrying phase with propagating solitary cluster waves. At the transition line between the two phases, the density is $\rhoc$.

In the dark green areas, the current increases linearly with the particle density. A nonlinear dependence can occur in the bright 
green areas, when $\rho$ exceeds $\rho_\ast=\rhoc+1/\mb^2$ and $\rho_\ast<1/\sigma$. We demonstrate
this nonlinear dependence in Fig.~\ref{fig:J-rho}(c).
In the inset of Fig.~\ref{fig:phase-diagrams-fixed-f}(a), we show the variation of $\nb$ with $\sigma$. Remarkably, this variation 
is rather irregular for $0.35\lesssim\sigma\lesssim0.8$. It gives rise to
dip-like changes of $\rhoc$, see e.g.\ the dips around $\sigma=0.5$, 0.67, and 0.75.

By contrast, at larger $f$, the behavior is more regular,
as demonstrated by the phase diagram in Fig.~\ref{fig:phase-diagrams-fixed-f}
(b) for $f=1$. For this larger drag force, $\nb$ decreases monotonically in a stepwise manner for $\sigma>0.05$ and $\mb$ is 
always one. According to Eq.~\eqref{eq:rhoc}, $\rhoc$ then decreases stepwise with $\sigma$.

Figure~\ref{fig:phase-diagrams-fixed-sigma} depicts a representative phase diagram in the $f\rho$-plane for $\sigma=3/25$. As $\nb$ is monotonically decreasing with $f$, a similar regular stepwise decrease of $\rhoc$ with $f$ occurs for all particle diameters.
Following a vertical line at $f=0.1$ in this diagram, one passes the transition point $\rhoc=7$ and the crossing point $\rho_\ast=8$ between linear and nonlinear parts of the current-density relation seen in Fig.~\ref{fig:J-rho}(c).

\subsection{Impact of phase transition on universality classes of current correlations}
\label{subsec:current-correlations}
In driven single-file transport, not only current-density relations are of interest but also 
correlations between thermal fluctuations of particle currents. In a frame comoving with the pattern velocity given by the derivative of the current,
\begin{equation}
v(\rho)=J'(\rho)\,,
\end{equation}
current correlation functions decay as power laws.
If $J''(\rho)\ne0$, exponents of these power laws belong to the Kardar-Parisi-Zhang (KPZ) universality class \cite{Beijeren/etal:1985, Ferrari/Spohn:2016, Spohn:2020}.
For $J''(\rho)=0$, by contrast, they are of the Edwards-Wilkinson (EW) universality class, as for purely diffusive
motion in the absence of a bias \cite{Krug:1997}.

\begin{figure}[b!]
\centering
\includegraphics[width=0.95\columnwidth]{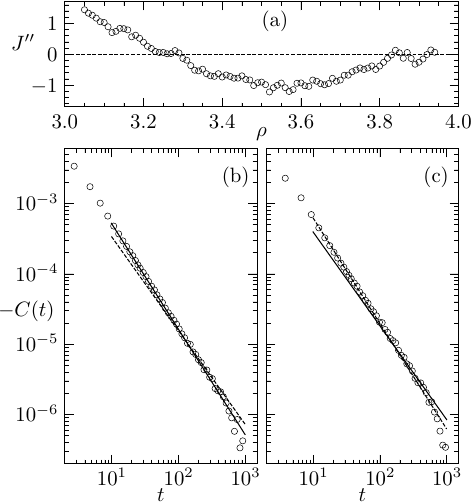}
\caption{\small Transition between KPZ and EW universality classes for correlations between current fluctuations. Parameters used are $\sigma=1/4$, $f=0.1$, and $D=0.05$.
 (a) Second derivative $J''(\rho)$ of the current-density relation shown in Fig.~\ref{fig:J-rho}(b). (b) Correlation function $C(t)$ of current density fluctuations [Eq.~\eqref{eq:C}] for $\rho=3.3$. The dashed and solid lines are least-squares fits
of power laws $C(t)=-C_\alpha t^{-\alpha}$ to the simulated data in the time range $10-10^3$
for fixed exponents $\alpha=4/3$ and $3/2$.  (c) Same as (b) for $\rho=3.05$, but with dashed and solid lines
corresponding to exponents $\alpha=3/2$ and 4/3.}
 \label{fig:current-correlations}
\end{figure}

Current-density relations are commonly nonlinear  for interacting particle systems. However, for $D=0$
solitary cluster waves 
give rise to a linear dependence of $J$ on $\rho$ above $\rho_{\rm c}$ 
according to Eq.~\eqref{eq:J-rho}.
One can understand this by viewing the solitons as independent quasiparticles moving with velocity $\vsol$.
In the presence of noise ($D>0$), the $J$-$\rho$-relation becomes nonlinear close to $\rho_{\rm c}$, because
soliton-mediated currents are weak and
thermal noise has a stronger influence.
Beyond this regime of weak soliton-mediated currents, a nearly linear $J$-$\rho$-relation is recovered.

Considering the current density $j(x,t)$, it has the form $j_{\rm cf}(x,t)=j(x+vt,t)-v\varrho(x+vt,t)$ in a comoving frame with velocity $v$, where  $\varrho(x,t)$ is the local particle density.
Fluctuations $\delta j_{\rm cf}(x,t)=j_{\rm cf}(x,t)-\langle j_{\rm cf}(x,t)\rangle$ are expected to be anticorrelated at long times and their
correlation function 
\begin{align}
C(t)= \langle\delta j_{\rm cf}(x,t)\delta j_{\rm cf}(x,0)\rangle
\label{eq:C}
\end{align}
to decay as $-t^{-4/3}$ in the KPZ and as $-t^{-3/2}$ in the EW case.

For calculating $C(t)$, we applied the method described in \cite{Schweers/etal:2025}. 
These calculations are demanding, taking computing times of ten days and more on 500 cores of
AMD EPYC 7452 processors for obtaining accurate results.
We analyze $C(t)$ at different densities $\rho>\rho_{\rm c}$ 
for the parameters
$D=0.05$, $\sigma=1/4$, and $f=0.1$, where $J(\rho)$ is given in Fig.~\ref{fig:J-rho}(b).
To calculate $J''(\rho)$ from this current-density relation, we fitted 
a quadratic polynomial to $J(\rho)$ in small $\rho$-intervals of size 0.1. The results 
are given in Fig.~\ref{fig:current-correlations}(a).

Around $\rho\cong3.3$, $J''(\rho)$ is 
negligibly small, and least-squares fits of the decay of $C(t)$ to power laws according to the KPZ (dashed line) and EW (solid line) scaling clearly signals a better match to the EW class, i.e.\ $C(t)\sim -t^{-3/2}$, see Fig.~\ref{fig:current-correlations}(b).
For $\rho=3.05$ by contrast, where $J''(\rho)\simeq1.4$, we obtain a better match to the KPZ scaling $C(t)\sim -t^{-4/3}$ (solid line), see Fig.~\ref{fig:current-correlations}(c).

Hence, the phase transition at $\rho_{\rm c}$ is reflected also in the correlation behavior of currents:
a change from the KPZ to the EW universality class occurs due to soliton-mediated transport.

\section{Summary and Outlook}

We have demonstrated a nonequilibrium phase transition  in a homogeneous single-file system of Brownian particles driven across highly crowded periodic structures. The transition is between a weak-current phase of thermally activated motion and a high-current phase of nonactivated particle transport due to formation of solitary cluster waves. 
It occurs at low noise at a critical particle density $\rhoc$ that is determined by the largest cluster stabilizable in the periodic energy landscape. The size of this cluster depends on the particle diameter and drag force acting on the particles, yielding a complex phase diagram.

Above $\rhoc$, solitons govern the particle transport. They let the current-density relation to become nearly linear since the number of solitons increases proportionally to the density and 
the solitary waves propagate with an almost constant velocity.
In a frame comoving with this velocity, fluctuations are due to unbiased diffusion of solitons as in an equilibrium. 
An EW scaling of current correlation occurs rather than the typical KPZ scaling in driven single-file transport. 
The phase transition is therefore reflected also in the behavior of current correlations.

Our quantitative description of the phase transition is based on the unit displacement law and further
results derived in Ref.~\cite{Antonov/etal:2025b}. There we have shown that the relevant features are present also for a
triangle wave potential, i.e.\ a potential very much different from the sinusoidal one. This strongly suggests that the phase transition occurs generally in periodic potentials at high crowding.

Solitary cluster waves have recently been observed in experiments with colloidal particles \cite{Cereceda-Lopez/etal:2023}, and we expect the phase transition reported here to be seen in corresponding measurements. It can play an important role
for biological transport in highly crowded periodic environments. In particular, the solitary waves lead to remarkably high particle currents in potential landscapes with barriers much larger than the thermal energy.

From the theoretical perspective, a challenging task is to develop
a continuum description of the cluster waves and the phase transition. A promising approach might be 
based on dynamic density \cite{Loewen:2017, teVrugt/etal:2020} or power functional \cite{Schmidt:2022} theory, or on macroscopic fluctuation theory \cite{Berlioz/etal:2025}.
An alternative could be to adapt a recently suggested
hydrodynamic field theory  \cite{Hurtado-Gutierrez/etal:2020, Hurtado-Gutierrez/etal:2025}
with a properly chosen packing field, as used to model time crystals.

\vspace{4ex}
\begin{acknowledgments} 
We thank P.~Tierno for very helpful discussions on soliton dynamics in experiments. We gratefully acknowledge financial support by the Czech Science Foundation (Project No.\ 23-09074L) and the Deutsche Forschungsgemeinschaft (Project No.\ 521001072), and the use of a high-performance computing cluster funded by the Deutsche Forschungsgemeinschaft (Project No.\ 456666331).
\end{acknowledgments}


\begin{thebibliography}{47}%
\makeatletter
\providecommand \@ifxundefined [1]{%
 \@ifx{#1\undefined}
}%
\providecommand \@ifnum [1]{%
 \ifnum #1\expandafter \@firstoftwo
 \else \expandafter \@secondoftwo
 \fi
}%
\providecommand \@ifx [1]{%
 \ifx #1\expandafter \@firstoftwo
 \else \expandafter \@secondoftwo
 \fi
}%
\providecommand \natexlab [1]{#1}%
\providecommand \enquote  [1]{``#1''}%
\providecommand \bibnamefont  [1]{#1}%
\providecommand \bibfnamefont [1]{#1}%
\providecommand \citenamefont [1]{#1}%
\providecommand \href@noop [0]{\@secondoftwo}%
\providecommand \href [0]{\begingroup \@sanitize@url \@href}%
\providecommand \@href[1]{\@@startlink{#1}\@@href}%
\providecommand \@@href[1]{\endgroup#1\@@endlink}%
\providecommand \@sanitize@url [0]{\catcode `\\12\catcode `\$12\catcode
  `\&12\catcode `\#12\catcode `\^12\catcode `\_12\catcode `\%12\relax}%
\providecommand \@@startlink[1]{}%
\providecommand \@@endlink[0]{}%
\providecommand \url  [0]{\begingroup\@sanitize@url \@url }%
\providecommand \@url [1]{\endgroup\@href {#1}{\urlprefix }}%
\providecommand \urlprefix  [0]{URL }%
\providecommand \Eprint [0]{\href }%
\providecommand \doibase [0]{https://doi.org/}%
\providecommand \selectlanguage [0]{\@gobble}%
\providecommand \bibinfo  [0]{\@secondoftwo}%
\providecommand \bibfield  [0]{\@secondoftwo}%
\providecommand \translation [1]{[#1]}%
\providecommand \BibitemOpen [0]{}%
\providecommand \bibitemStop [0]{}%
\providecommand \bibitemNoStop [0]{.\EOS\space}%
\providecommand \EOS [0]{\spacefactor3000\relax}%
\providecommand \BibitemShut  [1]{\csname bibitem#1\endcsname}%
\let\auto@bib@innerbib\@empty
\bibitem [{\citenamefont {Wei}\ \emph {et~al.}(2000)\citenamefont {Wei},
  \citenamefont {Bechinger},\ and\ \citenamefont {Leiderer}}]{Wei/etal:2000}%
  \BibitemOpen
  \bibfield  {author} {\bibinfo {author} {\bibfnamefont {Q.-H.}\ \bibnamefont
  {Wei}}, \bibinfo {author} {\bibfnamefont {C.}~\bibnamefont {Bechinger}},\
  and\ \bibinfo {author} {\bibfnamefont {P.}~\bibnamefont {Leiderer}},\
  }\bibfield  {title} {\bibinfo {title} {Single-file diffusion of colloids in
  one-dimensional channels},\ }\href
  {https://doi.org/10.1126/science.287.5453.625} {\bibfield  {journal}
  {\bibinfo  {journal} {Science}\ }\textbf {\bibinfo {volume} {287}},\ \bibinfo
  {pages} {625} (\bibinfo {year} {2000})}\BibitemShut {NoStop}%
\bibitem [{\citenamefont {Ma}\ \emph {et~al.}(2015)\citenamefont {Ma},
  \citenamefont {Grey}, \citenamefont {Shen}, \citenamefont {Urbakh},
  \citenamefont {Wu}, \citenamefont {Liu}, \citenamefont {Liu},\ and\
  \citenamefont {Zheng}}]{Ma/etal:2015}%
  \BibitemOpen
  \bibfield  {author} {\bibinfo {author} {\bibfnamefont {M.}~\bibnamefont
  {Ma}}, \bibinfo {author} {\bibfnamefont {F.}~\bibnamefont {Grey}}, \bibinfo
  {author} {\bibfnamefont {L.}~\bibnamefont {Shen}}, \bibinfo {author}
  {\bibfnamefont {M.}~\bibnamefont {Urbakh}}, \bibinfo {author} {\bibfnamefont
  {S.}~\bibnamefont {Wu}}, \bibinfo {author} {\bibfnamefont {J.~Z.}\
  \bibnamefont {Liu}}, \bibinfo {author} {\bibfnamefont {Y.}~\bibnamefont
  {Liu}},\ and\ \bibinfo {author} {\bibfnamefont {Q.}~\bibnamefont {Zheng}},\
  }\bibfield  {title} {\bibinfo {title} {Water transport inside carbon
  nanotubes mediated by phonon-induced oscillating friction},\ }\href
  {https://doi.org/10.1038/nnano.2015.134} {\bibfield  {journal} {\bibinfo
  {journal} {Nat. Nanotechnol.}\ }\textbf {\bibinfo {volume} {10}},\ \bibinfo
  {pages} {692} (\bibinfo {year} {2015})}\BibitemShut {NoStop}%
\bibitem [{\citenamefont {Wang}\ \emph {et~al.}(2024)\citenamefont {Wang},
  \citenamefont {Lu},\ and\ \citenamefont {Chen}}]{Wang/etal:2024}%
  \BibitemOpen
  \bibfield  {author} {\bibinfo {author} {\bibfnamefont {Y.}~\bibnamefont
  {Wang}}, \bibinfo {author} {\bibfnamefont {Y.}~\bibnamefont {Lu}},\ and\
  \bibinfo {author} {\bibfnamefont {J.}~\bibnamefont {Chen}},\ }\bibfield
  {title} {\bibinfo {title} {Spontaneous spatiotemporal ordering of single-file
  water flow in narrow nanotubes},\ }\href
  {https://doi.org/10.1021/acs.jpcc.4c01214} {\bibfield  {journal} {\bibinfo
  {journal} {J. Phys. Chem. C}\ }\textbf {\bibinfo {volume} {128}},\ \bibinfo
  {pages} {12488} (\bibinfo {year} {2024})}\BibitemShut {NoStop}%
\bibitem [{\citenamefont {Yiu}\ \emph {et~al.}(2001)\citenamefont {Yiu},
  \citenamefont {Botting}, \citenamefont {Botting},\ and\ \citenamefont
  {Wright}}]{Yiu/etal:2001}%
  \BibitemOpen
  \bibfield  {author} {\bibinfo {author} {\bibfnamefont {H.~H.~P.}\
  \bibnamefont {Yiu}}, \bibinfo {author} {\bibfnamefont {C.~H.}\ \bibnamefont
  {Botting}}, \bibinfo {author} {\bibfnamefont {N.~P.}\ \bibnamefont
  {Botting}},\ and\ \bibinfo {author} {\bibfnamefont {P.~A.}\ \bibnamefont
  {Wright}},\ }\bibfield  {title} {\bibinfo {title} {Size selective protein
  adsorption on thiol-functionalised {SBA-15} mesoporous molecular sieve},\
  }\href {https://doi.org/10.1039/B104729A} {\bibfield  {journal} {\bibinfo
  {journal} {Phys. Chem. Chem. Phys.}\ }\textbf {\bibinfo {volume} {3}},\
  \bibinfo {pages} {2983} (\bibinfo {year} {2001})}\BibitemShut {NoStop}%
\bibitem [{\citenamefont {Hartmann}(2005)}]{Hartmann:2005}%
  \BibitemOpen
  \bibfield  {author} {\bibinfo {author} {\bibfnamefont {M.}~\bibnamefont
  {Hartmann}},\ }\bibfield  {title} {\bibinfo {title} {Ordered mesoporous
  materials for bioadsorption and biocatalysis},\ }\href
  {https://doi.org/10.1021/cm0485658} {\bibfield  {journal} {\bibinfo
  {journal} {Chem. Mater.}\ }\textbf {\bibinfo {volume} {17}},\ \bibinfo
  {pages} {4577} (\bibinfo {year} {2005})}\BibitemShut {NoStop}%
\bibitem [{\citenamefont {Zeng}\ \emph {et~al.}(2018)\citenamefont {Zeng},
  \citenamefont {Chen}, \citenamefont {Wang}, \citenamefont {Zhou},
  \citenamefont {Chen},\ and\ \citenamefont {Dai}}]{Zeng/etal:2018}%
  \BibitemOpen
  \bibfield  {author} {\bibinfo {author} {\bibfnamefont {S.}~\bibnamefont
  {Zeng}}, \bibinfo {author} {\bibfnamefont {J.}~\bibnamefont {Chen}}, \bibinfo
  {author} {\bibfnamefont {X.}~\bibnamefont {Wang}}, \bibinfo {author}
  {\bibfnamefont {G.}~\bibnamefont {Zhou}}, \bibinfo {author} {\bibfnamefont
  {L.}~\bibnamefont {Chen}},\ and\ \bibinfo {author} {\bibfnamefont
  {C.}~\bibnamefont {Dai}},\ }\bibfield  {title} {\bibinfo {title} {Selective
  transport through the ultrashort carbon nanotubes embedded in lipid
  bilayers},\ }\href {https://doi.org/10.1021/acs.jpcc.8b07861} {\bibfield
  {journal} {\bibinfo  {journal} {J. Phys. Chem. C}\ }\textbf {\bibinfo
  {volume} {122}},\ \bibinfo {pages} {27681} (\bibinfo {year}
  {2018})}\BibitemShut {NoStop}%
\bibitem [{\citenamefont {Li}\ and\ \citenamefont {Noy}(2025)}]{Li/Noy:2025}%
  \BibitemOpen
  \bibfield  {author} {\bibinfo {author} {\bibfnamefont {Z.}~\bibnamefont
  {Li}}\ and\ \bibinfo {author} {\bibfnamefont {A.}~\bibnamefont {Noy}},\
  }\bibfield  {title} {\bibinfo {title} {Carbon nanotube nanofluidics},\
  }\bibfield  {journal} {\bibinfo  {journal} {Chem. Soc. Rev.}\ }\href
  {https://doi.org/10.1039/D5CS00233H} {10.1039/D5CS00233H} (\bibinfo {year}
  {2025})\BibitemShut {NoStop}%
\bibitem [{\citenamefont {Hahn}\ \emph {et~al.}(1996)\citenamefont {Hahn},
  \citenamefont {K\"arger},\ and\ \citenamefont {Kukla}}]{Hahn/etal:1996}%
  \BibitemOpen
  \bibfield  {author} {\bibinfo {author} {\bibfnamefont {K.}~\bibnamefont
  {Hahn}}, \bibinfo {author} {\bibfnamefont {J.}~\bibnamefont {K\"arger}},\
  and\ \bibinfo {author} {\bibfnamefont {V.}~\bibnamefont {Kukla}},\ }\bibfield
   {title} {\bibinfo {title} {Single-file diffusion observation},\ }\href
  {https://doi.org/10.1103/PhysRevLett.76.2762} {\bibfield  {journal} {\bibinfo
   {journal} {Phys. Rev. Lett.}\ }\textbf {\bibinfo {volume} {76}},\ \bibinfo
  {pages} {2762} (\bibinfo {year} {1996})}\BibitemShut {NoStop}%
\bibitem [{\citenamefont {{Van de Voorde}}\ and\ \citenamefont
  {Sels}(2017)}]{VanDeVoorde/Sels:2017}%
  \BibitemOpen
  \bibinfo {editor} {\bibfnamefont {M.}~\bibnamefont {{Van de Voorde}}}\ and\
  \bibinfo {editor} {\bibfnamefont {B.}~\bibnamefont {Sels}},\ eds.,\ \href
  {https://www.wiley.com/en-us/Nanotechnology+in+Catalysis%3A+Applications+in+the+Chemical+Industry%2C+Energy+Development%2C+and+Environment+Protection%2C+3+Volumes-p-9783527699827}
  {\emph {\bibinfo {title} {Nanotechnology in Catalysis: Applications in the
  Chemical Industry, Energy Development, and Environment Protection}}}\
  (\bibinfo  {publisher} {Wiley-VCH, Weinheim},\ \bibinfo {year}
  {2017})\BibitemShut {NoStop}%
\bibitem [{\citenamefont {Chmelik}\ \emph {et~al.}(2023)\citenamefont
  {Chmelik}, \citenamefont {Caro}, \citenamefont {Freude}, \citenamefont
  {Haase}, \citenamefont {Valiullin},\ and\ \citenamefont
  {K{\"a}rger}}]{Chmelik/etal:2023}%
  \BibitemOpen
  \bibfield  {author} {\bibinfo {author} {\bibfnamefont {C.}~\bibnamefont
  {Chmelik}}, \bibinfo {author} {\bibfnamefont {J.}~\bibnamefont {Caro}},
  \bibinfo {author} {\bibfnamefont {D.}~\bibnamefont {Freude}}, \bibinfo
  {author} {\bibfnamefont {J.}~\bibnamefont {Haase}}, \bibinfo {author}
  {\bibfnamefont {R.}~\bibnamefont {Valiullin}},\ and\ \bibinfo {author}
  {\bibfnamefont {J.}~\bibnamefont {K{\"a}rger}},\ }\bibinfo {title} {Diffusive
  spreading of molecules in nanoporous materials},\ in\ \href
  {https://doi.org/10.1007/978-3-031-05946-9_10} {\emph {\bibinfo {booktitle}
  {Diffusive Spreading in Nature, Technology and Society}}},\ \bibinfo {editor}
  {edited by\ \bibinfo {editor} {\bibfnamefont {A.}~\bibnamefont {Bunde}},
  \bibinfo {editor} {\bibfnamefont {J.}~\bibnamefont {Caro}}, \bibinfo {editor}
  {\bibfnamefont {C.}~\bibnamefont {Chmelik}}, \bibinfo {editor} {\bibfnamefont
  {J.}~\bibnamefont {K{\"a}rger}},\ and\ \bibinfo {editor} {\bibfnamefont
  {G.}~\bibnamefont {Vogl}}}\ (\bibinfo  {publisher} {Springer International
  Publishing},\ \bibinfo {address} {Cham},\ \bibinfo {year} {2023})\
  Chap.~\bibinfo {chapter} {10}, pp.\ \bibinfo {pages} {179--214}\BibitemShut
  {NoStop}%
\bibitem [{\citenamefont {Liu}\ \emph {et~al.}(2025)\citenamefont {Liu},
  \citenamefont {Kan}, \citenamefont {Gao}, \citenamefont {Ji}, \citenamefont
  {Ye}, \citenamefont {Tan}, \citenamefont {Liu}, \citenamefont {Yuan},
  \citenamefont {Tang}, \citenamefont {Li}, \citenamefont {Gao}, \citenamefont
  {Xue}, \citenamefont {Cai}, \citenamefont {Osti}, \citenamefont {Jalarvo},
  \citenamefont {Li}, \citenamefont {Zou}, \citenamefont {Li}, \citenamefont
  {Xu}, \citenamefont {Hou}, \citenamefont {Ye}, \citenamefont {Liu},\ and\
  \citenamefont {Zheng}}]{Liu/etal:2025}%
  \BibitemOpen
  \bibfield  {author} {\bibinfo {author} {\bibfnamefont {Z.}~\bibnamefont
  {Liu}}, \bibinfo {author} {\bibfnamefont {X.}~\bibnamefont {Kan}}, \bibinfo
  {author} {\bibfnamefont {M.}~\bibnamefont {Gao}}, \bibinfo {author}
  {\bibfnamefont {Y.}~\bibnamefont {Ji}}, \bibinfo {author} {\bibfnamefont
  {F.}~\bibnamefont {Ye}}, \bibinfo {author} {\bibfnamefont {J.}~\bibnamefont
  {Tan}}, \bibinfo {author} {\bibfnamefont {F.}~\bibnamefont {Liu}}, \bibinfo
  {author} {\bibfnamefont {J.}~\bibnamefont {Yuan}}, \bibinfo {author}
  {\bibfnamefont {X.}~\bibnamefont {Tang}}, \bibinfo {author} {\bibfnamefont
  {H.}~\bibnamefont {Li}}, \bibinfo {author} {\bibfnamefont {P.}~\bibnamefont
  {Gao}}, \bibinfo {author} {\bibfnamefont {J.}~\bibnamefont {Xue}}, \bibinfo
  {author} {\bibfnamefont {Q.}~\bibnamefont {Cai}}, \bibinfo {author}
  {\bibfnamefont {N.~C.}\ \bibnamefont {Osti}}, \bibinfo {author}
  {\bibfnamefont {N.~H.}\ \bibnamefont {Jalarvo}}, \bibinfo {author}
  {\bibfnamefont {C.}~\bibnamefont {Li}}, \bibinfo {author} {\bibfnamefont
  {Y.}~\bibnamefont {Zou}}, \bibinfo {author} {\bibfnamefont {Y.}~\bibnamefont
  {Li}}, \bibinfo {author} {\bibfnamefont {S.}~\bibnamefont {Xu}}, \bibinfo
  {author} {\bibfnamefont {G.}~\bibnamefont {Hou}}, \bibinfo {author}
  {\bibfnamefont {M.}~\bibnamefont {Ye}}, \bibinfo {author} {\bibfnamefont
  {F.}~\bibnamefont {Liu}},\ and\ \bibinfo {author} {\bibfnamefont
  {A.}~\bibnamefont {Zheng}},\ }\bibfield  {title} {\bibinfo {title}
  {Asymmetric rotations slow down diffusion under confinement},\ }\href
  {https://doi.org/10.1038/s41467-025-57242-6} {\bibfield  {journal} {\bibinfo
  {journal} {Nat. Commun.}\ }\textbf {\bibinfo {volume} {16}},\ \bibinfo
  {pages} {2018} (\bibinfo {year} {2025})}\BibitemShut {NoStop}%
\bibitem [{\citenamefont {Hille}(2001)}]{Hille:2001}%
  \BibitemOpen
  \bibfield  {author} {\bibinfo {author} {\bibfnamefont {B.}~\bibnamefont
  {Hille}},\ }\href@noop {} {\emph {\bibinfo {title} {Ion channels of excitable
  membranes}}},\ \bibinfo {edition} {3rd}\ ed.\ (\bibinfo  {publisher}
  {Sinauer},\ \bibinfo {address} {Sunderland, Mass.},\ \bibinfo {year}
  {2001})\BibitemShut {NoStop}%
\bibitem [{\citenamefont {Kolomeisky}(2013)}]{Kolomeisky:2013}%
  \BibitemOpen
  \bibfield  {author} {\bibinfo {author} {\bibfnamefont {A.~B.}\ \bibnamefont
  {Kolomeisky}},\ }\bibfield  {title} {\bibinfo {title} {Motor proteins and
  molecular motors: how to operate machines at the nanoscale},\ }\href
  {https://doi.org/10.1088/0953-8984/25/46/463101} {\bibfield  {journal}
  {\bibinfo  {journal} {J. Phys.: Condens. Matter}\ }\textbf {\bibinfo {volume}
  {25}},\ \bibinfo {pages} {463101} (\bibinfo {year} {2013})}\BibitemShut
  {NoStop}%
\bibitem [{\citenamefont {Ciandrini}\ \emph {et~al.}(2014)\citenamefont
  {Ciandrini}, \citenamefont {Romano},\ and\ \citenamefont
  {Parmeggiani}}]{Ciandrini/etal:2014}%
  \BibitemOpen
  \bibfield  {author} {\bibinfo {author} {\bibfnamefont {L.}~\bibnamefont
  {Ciandrini}}, \bibinfo {author} {\bibfnamefont {M.~C.}\ \bibnamefont
  {Romano}},\ and\ \bibinfo {author} {\bibfnamefont {A.}~\bibnamefont
  {Parmeggiani}},\ }\bibfield  {title} {\bibinfo {title} {Stepping and crowding
  of molecular motors: Statistical kinetics from an exclusion process
  perspective},\ }\href {https://doi.org/10.1016/j.bpj.2014.07.012} {\bibfield
  {journal} {\bibinfo  {journal} {Biophys. J.}\ }\textbf {\bibinfo {volume}
  {107}},\ \bibinfo {pages} {1176} (\bibinfo {year} {2014})}\BibitemShut
  {NoStop}%
\bibitem [{\citenamefont {Kolomeisky}(2015)}]{Kolomeisky:2015}%
  \BibitemOpen
  \bibfield  {author} {\bibinfo {author} {\bibfnamefont {A.~B.}\ \bibnamefont
  {Kolomeisky}},\ }\href {https://doi.org/10.1201/b18426} {\emph {\bibinfo
  {title} {Motor Proteins and Molecular Motors}}}\ (\bibinfo  {publisher} {CRC
  Press, Boca Raton},\ \bibinfo {year} {2015})\BibitemShut {NoStop}%
\bibitem [{\citenamefont {Ariga}(2024)}]{Ariga:2024}%
  \BibitemOpen
  \bibfield  {author} {\bibinfo {author} {\bibfnamefont {T.}~\bibnamefont
  {Ariga}},\ }\bibfield  {title} {\bibinfo {title} {Nonthermal fluctuations
  accelerate biomolecular motors},\ }\href
  {https://doi.org/10.1007/s12551-024-01238-x} {\bibfield  {journal} {\bibinfo
  {journal} {Biophys. Rev.}\ }\textbf {\bibinfo {volume} {16}},\ \bibinfo
  {pages} {605} (\bibinfo {year} {2024})}\BibitemShut {NoStop}%
\bibitem [{\citenamefont {Grieb}\ \emph {et~al.}(2025)\citenamefont {Grieb},
  \citenamefont {Krishnan},\ and\ \citenamefont {Ross}}]{Grieb/etal:2025}%
  \BibitemOpen
  \bibfield  {author} {\bibinfo {author} {\bibfnamefont {M.}~\bibnamefont
  {Grieb}}, \bibinfo {author} {\bibfnamefont {N.}~\bibnamefont {Krishnan}},\
  and\ \bibinfo {author} {\bibfnamefont {J.~L.}\ \bibnamefont {Ross}},\
  }\bibfield  {title} {\bibinfo {title} {Multimotor cargo navigation in
  microtubule networks with various mesh sizes},\ }\href
  {https://doi.org/10.1103/PhysRevE.111.024413} {\bibfield  {journal} {\bibinfo
   {journal} {Phys. Rev. E}\ }\textbf {\bibinfo {volume} {111}},\ \bibinfo
  {pages} {024413} (\bibinfo {year} {2025})}\BibitemShut {NoStop}%
\bibitem [{\citenamefont {MacDonald}\ \emph {et~al.}(1968)\citenamefont
  {MacDonald}, \citenamefont {Gibbs},\ and\ \citenamefont
  {Pipkin}}]{MacDonald/etal:1968}%
  \BibitemOpen
  \bibfield  {author} {\bibinfo {author} {\bibfnamefont {C.~T.}\ \bibnamefont
  {MacDonald}}, \bibinfo {author} {\bibfnamefont {J.~H.}\ \bibnamefont
  {Gibbs}},\ and\ \bibinfo {author} {\bibfnamefont {A.~C.}\ \bibnamefont
  {Pipkin}},\ }\bibfield  {title} {\bibinfo {title} {Kinetics of
  biopolymerization on nucleic acid templates},\ }\href
  {https://doi.org/10.1002/bip.1968.360060102} {\bibfield  {journal} {\bibinfo
  {journal} {Biopolymers}\ }\textbf {\bibinfo {volume} {6}},\ \bibinfo {pages}
  {1} (\bibinfo {year} {1968})}\BibitemShut {NoStop}%
\bibitem [{\citenamefont {Erdmann-Pham}\ \emph {et~al.}(2020)\citenamefont
  {Erdmann-Pham}, \citenamefont {Dao~Duc},\ and\ \citenamefont
  {Song}}]{Erdmann-Pham:2020}%
  \BibitemOpen
  \bibfield  {author} {\bibinfo {author} {\bibfnamefont {D.~D.}\ \bibnamefont
  {Erdmann-Pham}}, \bibinfo {author} {\bibfnamefont {K.}~\bibnamefont
  {Dao~Duc}},\ and\ \bibinfo {author} {\bibfnamefont {Y.~S.}\ \bibnamefont
  {Song}},\ }\bibfield  {title} {\bibinfo {title} {The key parameters that
  govern translation efficiency},\ }\href
  {https://doi.org/10.1016/j.cels.2019.12.003} {\bibfield  {journal} {\bibinfo
  {journal} {Cell Syst.}\ }\textbf {\bibinfo {volume} {10}},\ \bibinfo {pages}
  {183} (\bibinfo {year} {2020})}\BibitemShut {NoStop}%
\bibitem [{\citenamefont {Antonov}\ \emph
  {et~al.}(2022{\natexlab{a}})\citenamefont {Antonov}, \citenamefont {Ryabov},\
  and\ \citenamefont {Maass}}]{Antonov/etal:2022a}%
  \BibitemOpen
  \bibfield  {author} {\bibinfo {author} {\bibfnamefont {A.~P.}\ \bibnamefont
  {Antonov}}, \bibinfo {author} {\bibfnamefont {A.}~\bibnamefont {Ryabov}},\
  and\ \bibinfo {author} {\bibfnamefont {P.}~\bibnamefont {Maass}},\ }\bibfield
   {title} {\bibinfo {title} {Solitons in overdamped {B}rownian dynamics},\
  }\href {https://doi.org/10.1103/PhysRevLett.129.080601} {\bibfield  {journal}
  {\bibinfo  {journal} {Phys. Rev. Lett.}\ }\textbf {\bibinfo {volume} {129}},\
  \bibinfo {pages} {080601} (\bibinfo {year} {2022}{\natexlab{a}})}\BibitemShut
  {NoStop}%
\bibitem [{\citenamefont {Cereceda-L\'opez}\ \emph {et~al.}(2023)\citenamefont
  {Cereceda-L\'opez}, \citenamefont {Antonov}, \citenamefont {Ryabov},
  \citenamefont {Maass},\ and\ \citenamefont
  {Tierno}}]{Cereceda-Lopez/etal:2023}%
  \BibitemOpen
  \bibfield  {author} {\bibinfo {author} {\bibfnamefont {E.}~\bibnamefont
  {Cereceda-L\'opez}}, \bibinfo {author} {\bibfnamefont {A.~P.}\ \bibnamefont
  {Antonov}}, \bibinfo {author} {\bibfnamefont {A.}~\bibnamefont {Ryabov}},
  \bibinfo {author} {\bibfnamefont {P.}~\bibnamefont {Maass}},\ and\ \bibinfo
  {author} {\bibfnamefont {P.}~\bibnamefont {Tierno}},\ }\bibfield  {title}
  {\bibinfo {title} {Overcrowding induces fast colloidal solitons in a slowly
  rotating potential landscape},\ }\href
  {https://doi.org/10.1038/s41467-023-41989-x} {\bibfield  {journal} {\bibinfo
  {journal} {Nat. Commun.}\ }\textbf {\bibinfo {volume} {14}},\ \bibinfo
  {pages} {6448} (\bibinfo {year} {2023})}\BibitemShut {NoStop}%
\bibitem [{\citenamefont {Krug}(1991)}]{Krug:1991}%
  \BibitemOpen
  \bibfield  {author} {\bibinfo {author} {\bibfnamefont {J.}~\bibnamefont
  {Krug}},\ }\bibfield  {title} {\bibinfo {title} {Boundary-induced phase
  transitions in driven diffusive systems},\ }\href
  {https://doi.org/10.1103/PhysRevLett.67.1882} {\bibfield  {journal} {\bibinfo
   {journal} {Phys. Rev. Lett.}\ }\textbf {\bibinfo {volume} {67}},\ \bibinfo
  {pages} {1882} (\bibinfo {year} {1991})}\BibitemShut {NoStop}%
\bibitem [{\citenamefont {Sch{\"u}tz}\ and\ \citenamefont
  {Domany}(1993)}]{Schuetz/Domany:1993}%
  \BibitemOpen
  \bibfield  {author} {\bibinfo {author} {\bibfnamefont {G.}~\bibnamefont
  {Sch{\"u}tz}}\ and\ \bibinfo {author} {\bibfnamefont {E.}~\bibnamefont
  {Domany}},\ }\bibfield  {title} {\bibinfo {title} {Phase transitions in an
  exactly soluble one-dimensional exclusion process},\ }\href
  {https://doi.org/10.1007/BF01048050} {\bibfield  {journal} {\bibinfo
  {journal} {J. Stat. Phys.}\ }\textbf {\bibinfo {volume} {72}},\ \bibinfo
  {pages} {277} (\bibinfo {year} {1993})}\BibitemShut {NoStop}%
\bibitem [{\citenamefont {Kolomeisky}(1998)}]{Kolomeisky:1998}%
  \BibitemOpen
  \bibfield  {author} {\bibinfo {author} {\bibfnamefont {A.~B.}\ \bibnamefont
  {Kolomeisky}},\ }\bibfield  {title} {\bibinfo {title} {Asymmetric simple
  exclusion model with local inhomogeneity},\ }\href
  {https://doi.org/10.1088/0305-4470/31/4/006} {\bibfield  {journal} {\bibinfo
  {journal} {J. Phys. A: Math. Gen.}\ }\textbf {\bibinfo {volume} {31}},\
  \bibinfo {pages} {1153} (\bibinfo {year} {1998})}\BibitemShut {NoStop}%
\bibitem [{\citenamefont {Hinsch}\ and\ \citenamefont
  {Frey}(2006)}]{Hinsch/Frey:2006}%
  \BibitemOpen
  \bibfield  {author} {\bibinfo {author} {\bibfnamefont {H.}~\bibnamefont
  {Hinsch}}\ and\ \bibinfo {author} {\bibfnamefont {E.}~\bibnamefont {Frey}},\
  }\bibfield  {title} {\bibinfo {title} {Bulk-driven nonequilibrium phase
  transitions in a mesoscopic ring},\ }\href
  {https://doi.org/10.1103/PhysRevLett.97.095701} {\bibfield  {journal}
  {\bibinfo  {journal} {Phys. Rev. Lett.}\ }\textbf {\bibinfo {volume} {97}},\
  \bibinfo {pages} {095701} (\bibinfo {year} {2006})}\BibitemShut {NoStop}%
\bibitem [{\citenamefont {Brzank}\ and\ \citenamefont
  {Sch{\"u}tz}(2007)}]{Brzank/Schuetz:2007}%
  \BibitemOpen
  \bibfield  {author} {\bibinfo {author} {\bibfnamefont {A.}~\bibnamefont
  {Brzank}}\ and\ \bibinfo {author} {\bibfnamefont {G.~M.}\ \bibnamefont
  {Sch{\"u}tz}},\ }\bibfield  {title} {\bibinfo {title} {Phase transition in
  the two-component symmetric exclusion process with open boundaries},\ }\href
  {https://doi.org/10.1088/1742-5468/2007/08/p08028} {\bibfield  {journal}
  {\bibinfo  {journal} {J. Stat. Mech. Theor. Exp.}\ }\textbf {\bibinfo
  {volume} {2007}},\ \bibinfo {pages} {P08028} (\bibinfo {year}
  {2007})}\BibitemShut {NoStop}%
\bibitem [{\citenamefont {Bodineau}\ and\ \citenamefont
  {Derrida}(2005)}]{Bodineau/Derrida:2005}%
  \BibitemOpen
  \bibfield  {author} {\bibinfo {author} {\bibfnamefont {T.}~\bibnamefont
  {Bodineau}}\ and\ \bibinfo {author} {\bibfnamefont {B.}~\bibnamefont
  {Derrida}},\ }\bibfield  {title} {\bibinfo {title} {Distribution of current
  in nonequilibrium diffusive systems and phase transitions},\ }\href
  {https://doi.org/10.1103/PhysRevE.72.066110} {\bibfield  {journal} {\bibinfo
  {journal} {Phys. Rev. E}\ }\textbf {\bibinfo {volume} {72}},\ \bibinfo
  {pages} {066110} (\bibinfo {year} {2005})}\BibitemShut {NoStop}%
\bibitem [{\citenamefont {Lips}\ \emph {et~al.}(2018)\citenamefont {Lips},
  \citenamefont {Ryabov},\ and\ \citenamefont {Maass}}]{Lips/etal:2018}%
  \BibitemOpen
  \bibfield  {author} {\bibinfo {author} {\bibfnamefont {D.}~\bibnamefont
  {Lips}}, \bibinfo {author} {\bibfnamefont {A.}~\bibnamefont {Ryabov}},\ and\
  \bibinfo {author} {\bibfnamefont {P.}~\bibnamefont {Maass}},\ }\bibfield
  {title} {\bibinfo {title} {Brownian asymmetric simple exclusion process},\
  }\href {https://doi.org/10.1103/PhysRevLett.121.160601} {\bibfield  {journal}
  {\bibinfo  {journal} {Phys. Rev. Lett.}\ }\textbf {\bibinfo {volume} {121}},\
  \bibinfo {pages} {160601} (\bibinfo {year} {2018})}\BibitemShut {NoStop}%
\bibitem [{\citenamefont {Lips}\ \emph {et~al.}(2019)\citenamefont {Lips},
  \citenamefont {Ryabov},\ and\ \citenamefont {Maass}}]{Lips/etal:2019}%
  \BibitemOpen
  \bibfield  {author} {\bibinfo {author} {\bibfnamefont {D.}~\bibnamefont
  {Lips}}, \bibinfo {author} {\bibfnamefont {A.}~\bibnamefont {Ryabov}},\ and\
  \bibinfo {author} {\bibfnamefont {P.}~\bibnamefont {Maass}},\ }\bibfield
  {title} {\bibinfo {title} {Single-file transport in periodic potentials: {The
  Brownian} asymmetric simple exclusion process},\ }\href
  {https://doi.org/10.1103/PhysRevE.100.052121} {\bibfield  {journal} {\bibinfo
   {journal} {Phys. Rev. E}\ }\textbf {\bibinfo {volume} {100}},\ \bibinfo
  {pages} {052121} (\bibinfo {year} {2019})}\BibitemShut {NoStop}%
\bibitem [{\citenamefont {Ryabov}\ \emph {et~al.}(2019)\citenamefont {Ryabov},
  \citenamefont {Lips},\ and\ \citenamefont {Maass}}]{Ryabov/etal:2019}%
  \BibitemOpen
  \bibfield  {author} {\bibinfo {author} {\bibfnamefont {A.}~\bibnamefont
  {Ryabov}}, \bibinfo {author} {\bibfnamefont {D.}~\bibnamefont {Lips}},\ and\
  \bibinfo {author} {\bibfnamefont {P.}~\bibnamefont {Maass}},\ }\bibfield
  {title} {\bibinfo {title} {Counterintuitive short uphill transitions in
  single-file diffusion},\ }\href {https://doi.org/10.1021/acs.jpcc.8b12081}
  {\bibfield  {journal} {\bibinfo  {journal} {J. Phys. Chem. C}\ }\textbf
  {\bibinfo {volume} {123}},\ \bibinfo {pages} {5714} (\bibinfo {year}
  {2019})}\BibitemShut {NoStop}%
\bibitem [{\citenamefont {Cereceda-L\'opez}\ \emph {et~al.}(2021)\citenamefont
  {Cereceda-L\'opez}, \citenamefont {Lips}, \citenamefont {Ortiz-Ambriz},
  \citenamefont {Ryabov}, \citenamefont {Maass},\ and\ \citenamefont
  {Tierno}}]{Cereceda-Lopez/etal:2021}%
  \BibitemOpen
  \bibfield  {author} {\bibinfo {author} {\bibfnamefont {E.}~\bibnamefont
  {Cereceda-L\'opez}}, \bibinfo {author} {\bibfnamefont {D.}~\bibnamefont
  {Lips}}, \bibinfo {author} {\bibfnamefont {A.}~\bibnamefont {Ortiz-Ambriz}},
  \bibinfo {author} {\bibfnamefont {A.}~\bibnamefont {Ryabov}}, \bibinfo
  {author} {\bibfnamefont {P.}~\bibnamefont {Maass}},\ and\ \bibinfo {author}
  {\bibfnamefont {P.}~\bibnamefont {Tierno}},\ }\bibfield  {title} {\bibinfo
  {title} {Hydrodynamic interactions can induce jamming in flow-driven
  systems},\ }\href {https://doi.org/10.1103/PhysRevLett.127.214501} {\bibfield
   {journal} {\bibinfo  {journal} {Phys. Rev. Lett.}\ }\textbf {\bibinfo
  {volume} {127}},\ \bibinfo {pages} {214501} (\bibinfo {year}
  {2021})}\BibitemShut {NoStop}%
\bibitem [{\citenamefont {Antonov}\ \emph
  {et~al.}(2025{\natexlab{a}})\citenamefont {Antonov}, \citenamefont
  {Vonhusen}, \citenamefont {Ryabov},\ and\ \citenamefont
  {Maass}}]{Antonov/etal:2025b}%
  \BibitemOpen
  \bibfield  {author} {\bibinfo {author} {\bibfnamefont {A.~P.}\ \bibnamefont
  {Antonov}}, \bibinfo {author} {\bibfnamefont {A.}~\bibnamefont {Vonhusen}},
  \bibinfo {author} {\bibfnamefont {A.}~\bibnamefont {Ryabov}},\ and\ \bibinfo
  {author} {\bibfnamefont {P.}~\bibnamefont {Maass}},\ }\bibfield  {title}
  {\bibinfo {title} {Cluster sizes, particle displacements and currents in
  transport mediated by solitary cluster waves},\ }\href
  {https://doi.org/10.1007/s11071-025-11626-x} {\bibfield  {journal} {\bibinfo
  {journal} {Nonlinear Dyn.}\ }\textbf {\bibinfo {volume} {113}},\ \bibinfo
  {pages} {31529} (\bibinfo {year} {2025}{\natexlab{a}})}\BibitemShut {NoStop}%
\bibitem [{\citenamefont {Antonov}\ \emph {et~al.}(2024)\citenamefont
  {Antonov}, \citenamefont {Ryabov},\ and\ \citenamefont
  {Maass}}]{Antonov/etal:2024}%
  \BibitemOpen
  \bibfield  {author} {\bibinfo {author} {\bibfnamefont {A.~P.}\ \bibnamefont
  {Antonov}}, \bibinfo {author} {\bibfnamefont {A.}~\bibnamefont {Ryabov}},\
  and\ \bibinfo {author} {\bibfnamefont {P.}~\bibnamefont {Maass}},\ }\bibfield
   {title} {\bibinfo {title} {Solitary cluster waves in periodic potentials:
  Formation, propagation, and soliton-mediated particle transport},\ }\href
  {https://doi.org/10.1016/j.chaos.2024.115079} {\bibfield  {journal} {\bibinfo
   {journal} {Chaos, Solitons \& Fractals}\ }\textbf {\bibinfo {volume}
  {185}},\ \bibinfo {pages} {115079} (\bibinfo {year} {2024})}\BibitemShut
  {NoStop}%
\bibitem [{\citenamefont {Antonov}\ \emph
  {et~al.}(2022{\natexlab{b}})\citenamefont {Antonov}, \citenamefont
  {Schweers}, \citenamefont {Ryabov},\ and\ \citenamefont
  {Maass}}]{Antonov/etal:2022c}%
  \BibitemOpen
  \bibfield  {author} {\bibinfo {author} {\bibfnamefont {A.~P.}\ \bibnamefont
  {Antonov}}, \bibinfo {author} {\bibfnamefont {S.}~\bibnamefont {Schweers}},
  \bibinfo {author} {\bibfnamefont {A.}~\bibnamefont {Ryabov}},\ and\ \bibinfo
  {author} {\bibfnamefont {P.}~\bibnamefont {Maass}},\ }\bibfield  {title}
  {\bibinfo {title} {Brownian dynamics simulations of hard rods in external
  fields and with contact interactions},\ }\href
  {https://doi.org/10.1103/PhysRevE.106.054606} {\bibfield  {journal} {\bibinfo
   {journal} {Phys. Rev. E}\ }\textbf {\bibinfo {volume} {106}},\ \bibinfo
  {pages} {054606} (\bibinfo {year} {2022}{\natexlab{b}})}\BibitemShut
  {NoStop}%
\bibitem [{\citenamefont {Antonov}\ \emph
  {et~al.}(2025{\natexlab{b}})\citenamefont {Antonov}, \citenamefont
  {Schweers}, \citenamefont {Ryabov},\ and\ \citenamefont
  {Maass}}]{Antonov/etal:2025a}%
  \BibitemOpen
  \bibfield  {author} {\bibinfo {author} {\bibfnamefont {A.~P.}\ \bibnamefont
  {Antonov}}, \bibinfo {author} {\bibfnamefont {S.}~\bibnamefont {Schweers}},
  \bibinfo {author} {\bibfnamefont {A.}~\bibnamefont {Ryabov}},\ and\ \bibinfo
  {author} {\bibfnamefont {P.}~\bibnamefont {Maass}},\ }\bibfield  {title}
  {\bibinfo {title} {Fast {B}rownian cluster dynamics},\ }\href
  {https://doi.org/10.1016/j.cpc.2024.109474} {\bibfield  {journal} {\bibinfo
  {journal} {Comput. Phys. Commun.}\ }\textbf {\bibinfo {volume} {309}},\
  \bibinfo {pages} {109474} (\bibinfo {year} {2025}{\natexlab{b}})}\BibitemShut
  {NoStop}%
\bibitem [{\citenamefont {Mishra}\ \emph {et~al.}(2025)\citenamefont {Mishra},
  \citenamefont {Ryabov},\ and\ \citenamefont {Maass}}]{Mishra/etal:2025}%
  \BibitemOpen
  \bibfield  {author} {\bibinfo {author} {\bibfnamefont {S.}~\bibnamefont
  {Mishra}}, \bibinfo {author} {\bibfnamefont {A.}~\bibnamefont {Ryabov}},\
  and\ \bibinfo {author} {\bibfnamefont {P.}~\bibnamefont {Maass}},\ }\bibfield
   {title} {\bibinfo {title} {Phase locking and fractional {S}hapiro steps in
  collective dynamics of microparticles},\ }\href
  {https://doi.org/10.1103/PhysRevLett.134.107102} {\bibfield  {journal}
  {\bibinfo  {journal} {Phys. Rev. Lett.}\ }\textbf {\bibinfo {volume} {134}},\
  \bibinfo {pages} {107102} (\bibinfo {year} {2025})}\BibitemShut {NoStop}%
\bibitem [{\citenamefont {van Beijeren}\ \emph {et~al.}(1985)\citenamefont {van
  Beijeren}, \citenamefont {Kutner},\ and\ \citenamefont
  {Spohn}}]{Beijeren/etal:1985}%
  \BibitemOpen
  \bibfield  {author} {\bibinfo {author} {\bibfnamefont {H.}~\bibnamefont {van
  Beijeren}}, \bibinfo {author} {\bibfnamefont {R.}~\bibnamefont {Kutner}},\
  and\ \bibinfo {author} {\bibfnamefont {H.}~\bibnamefont {Spohn}},\ }\bibfield
   {title} {\bibinfo {title} {Excess noise for driven diffusive systems},\
  }\href {https://doi.org/10.1103/PhysRevLett.54.2026} {\bibfield  {journal}
  {\bibinfo  {journal} {Phys. Rev. Lett.}\ }\textbf {\bibinfo {volume} {54}},\
  \bibinfo {pages} {2026} (\bibinfo {year} {1985})}\BibitemShut {NoStop}%
\bibitem [{\citenamefont {Ferrari}\ and\ \citenamefont
  {Spohn}(2016)}]{Ferrari/Spohn:2016}%
  \BibitemOpen
  \bibfield  {author} {\bibinfo {author} {\bibfnamefont {P.~L.}\ \bibnamefont
  {Ferrari}}\ and\ \bibinfo {author} {\bibfnamefont {H.}~\bibnamefont
  {Spohn}},\ }\bibfield  {title} {\bibinfo {title} {On time correlations for
  {KPZ} growth in one dimension},\ }\href
  {https://doi.org/10.3842/SIGMA.2016.074} {\bibfield  {journal} {\bibinfo
  {journal} {SIGMA}\ }\textbf {\bibinfo {volume} {12}},\ \bibinfo {pages} {074}
  (\bibinfo {year} {2016})}\BibitemShut {NoStop}%
\bibitem [{\citenamefont {Spohn}(2020)}]{Spohn:2020}%
  \BibitemOpen
  \bibfield  {author} {\bibinfo {author} {\bibfnamefont {H.}~\bibnamefont
  {Spohn}},\ }\bibfield  {title} {\bibinfo {title} {The 1 + 1 dimensional
  {K}ardar--{P}arisi--{Z}hang equation: more surprises},\ }\href
  {https://doi.org/10.1088/1742-5468/ab712a} {\bibfield  {journal} {\bibinfo
  {journal} {J. Stat. Mech: Theory Exp.}\ }\textbf {\bibinfo {volume} {2020}},\
  \bibinfo {pages} {044001} (\bibinfo {year} {2020})}\BibitemShut {NoStop}%
\bibitem [{\citenamefont {Krug}(1997)}]{Krug:1997}%
  \BibitemOpen
  \bibfield  {author} {\bibinfo {author} {\bibfnamefont {J.}~\bibnamefont
  {Krug}},\ }\bibfield  {title} {\bibinfo {title} {Origins of scale invariance
  in growth processes},\ }\href {https://doi.org/10.1080/00018739700101498}
  {\bibfield  {journal} {\bibinfo  {journal} {Adv. Phys.}\ }\textbf {\bibinfo
  {volume} {46}},\ \bibinfo {pages} {139} (\bibinfo {year} {1997})}\BibitemShut
  {NoStop}%
\bibitem [{\citenamefont {Schweers}\ \emph {et~al.}(2025)\citenamefont
  {Schweers}, \citenamefont {Sch\"utz},\ and\ \citenamefont
  {Maass}}]{Schweers/etal:2025}%
  \BibitemOpen
  \bibfield  {author} {\bibinfo {author} {\bibfnamefont {S.}~\bibnamefont
  {Schweers}}, \bibinfo {author} {\bibfnamefont {G.~M.}\ \bibnamefont
  {Sch\"utz}},\ and\ \bibinfo {author} {\bibfnamefont {P.}~\bibnamefont
  {Maass}},\ }\bibfield  {title} {\bibinfo {title} {Correlations of density and
  current fluctuations in single-file motion of hard spheres and in driven
  lattice gas with nearest-neighbor interaction},\ }\href
  {https://doi.org/10.1063/5.0266744} {\bibfield  {journal} {\bibinfo
  {journal} {J. Chem. Phys.}\ }\textbf {\bibinfo {volume} {162}},\ \bibinfo
  {pages} {164110} (\bibinfo {year} {2025})}\BibitemShut {NoStop}%
\bibitem [{\citenamefont {L\"owen}(2017)}]{Loewen:2017}%
  \BibitemOpen
  \bibfield  {author} {\bibinfo {author} {\bibfnamefont {H.}~\bibnamefont
  {L\"owen}},\ }\bibinfo {title} {Dynamical density functional theory for
  {B}rownian dynamics of colloidal particles},\ in\ \href
  {https://doi.org/10.1007/978-981-10-2502-0_9} {\emph {\bibinfo {booktitle}
  {Variational Methods in Molecular Modeling}}}\ (\bibinfo  {publisher}
  {Springer Singapore},\ \bibinfo {address} {Singapore},\ \bibinfo {year}
  {2017})\ pp.\ \bibinfo {pages} {255--284}\BibitemShut {NoStop}%
\bibitem [{\citenamefont {te~Vrugt}\ \emph {et~al.}(2020)\citenamefont
  {te~Vrugt}, \citenamefont {L{\"o}wen},\ and\ \citenamefont
  {Wittkowski}}]{teVrugt/etal:2020}%
  \BibitemOpen
  \bibfield  {author} {\bibinfo {author} {\bibfnamefont {M.}~\bibnamefont
  {te~Vrugt}}, \bibinfo {author} {\bibfnamefont {H.}~\bibnamefont
  {L{\"o}wen}},\ and\ \bibinfo {author} {\bibfnamefont {R.}~\bibnamefont
  {Wittkowski}},\ }\bibfield  {title} {\bibinfo {title} {Classical dynamical
  density functional theory: from fundamentals to applications},\ }\href
  {https://doi.org/10.1080/00018732.2020.1854965} {\bibfield  {journal}
  {\bibinfo  {journal} {Adv. Phys.}\ }\textbf {\bibinfo {volume} {69}},\
  \bibinfo {pages} {121} (\bibinfo {year} {2020})}\BibitemShut {NoStop}%
\bibitem [{\citenamefont {Schmidt}(2022)}]{Schmidt:2022}%
  \BibitemOpen
  \bibfield  {author} {\bibinfo {author} {\bibfnamefont {M.}~\bibnamefont
  {Schmidt}},\ }\bibfield  {title} {\bibinfo {title} {Power functional theory
  for many-body dynamics},\ }\href
  {https://doi.org/10.1103/RevModPhys.94.015007} {\bibfield  {journal}
  {\bibinfo  {journal} {Rev. Mod. Phys.}\ }\textbf {\bibinfo {volume} {94}},\
  \bibinfo {pages} {015007} (\bibinfo {year} {2022})}\BibitemShut {NoStop}%
\bibitem [{\citenamefont {Berlioz}\ \emph {et~al.}(2025)\citenamefont
  {Berlioz}, \citenamefont {B\'enichou},\ and\ \citenamefont
  {Grabsch}}]{Berlioz/etal:2025}%
  \BibitemOpen
  \bibfield  {author} {\bibinfo {author} {\bibfnamefont {T.}~\bibnamefont
  {Berlioz}}, \bibinfo {author} {\bibfnamefont {O.}~\bibnamefont
  {B\'enichou}},\ and\ \bibinfo {author} {\bibfnamefont {A.}~\bibnamefont
  {Grabsch}},\ }\bibfield  {title} {\bibinfo {title} {Tracer and current
  fluctuations in driven diffusive systems},\ }\href
  {https://doi.org/10.1103/4j5q-j4ht} {\bibfield  {journal} {\bibinfo
  {journal} {Phys. Rev. Lett.}\ }\textbf {\bibinfo {volume} {134}},\ \bibinfo
  {pages} {247101} (\bibinfo {year} {2025})}\BibitemShut {NoStop}%
\bibitem [{\citenamefont {Hurtado-Guti\'errez}\ \emph
  {et~al.}(2020)\citenamefont {Hurtado-Guti\'errez}, \citenamefont {Carollo},
  \citenamefont {P\'erez-Espigares},\ and\ \citenamefont
  {Hurtado}}]{Hurtado-Gutierrez/etal:2020}%
  \BibitemOpen
  \bibfield  {author} {\bibinfo {author} {\bibfnamefont {R.}~\bibnamefont
  {Hurtado-Guti\'errez}}, \bibinfo {author} {\bibfnamefont {F.}~\bibnamefont
  {Carollo}}, \bibinfo {author} {\bibfnamefont {C.}~\bibnamefont
  {P\'erez-Espigares}},\ and\ \bibinfo {author} {\bibfnamefont {P.~I.}\
  \bibnamefont {Hurtado}},\ }\bibfield  {title} {\bibinfo {title} {Building
  continuous time crystals from rare events},\ }\href
  {https://doi.org/10.1103/PhysRevLett.125.160601} {\bibfield  {journal}
  {\bibinfo  {journal} {Phys. Rev. Lett.}\ }\textbf {\bibinfo {volume} {125}},\
  \bibinfo {pages} {160601} (\bibinfo {year} {2020})}\BibitemShut {NoStop}%
\bibitem [{\citenamefont {Hurtado-Guti\'errez}\ \emph
  {et~al.}(2025)\citenamefont {Hurtado-Guti\'errez}, \citenamefont
  {P\'erez-Espigares},\ and\ \citenamefont
  {Hurtado}}]{Hurtado-Gutierrez/etal:2025}%
  \BibitemOpen
  \bibfield  {author} {\bibinfo {author} {\bibfnamefont {R.}~\bibnamefont
  {Hurtado-Guti\'errez}}, \bibinfo {author} {\bibfnamefont {C.}~\bibnamefont
  {P\'erez-Espigares}},\ and\ \bibinfo {author} {\bibfnamefont {P.~I.}\
  \bibnamefont {Hurtado}},\ }\bibfield  {title} {\bibinfo {title} {Programmable
  time crystals from higher-order packing fields},\ }\href
  {https://doi.org/10.1103/PhysRevE.111.034119} {\bibfield  {journal} {\bibinfo
   {journal} {Phys. Rev. E}\ }\textbf {\bibinfo {volume} {111}},\ \bibinfo
  {pages} {034119} (\bibinfo {year} {2025})}\BibitemShut {NoStop}%
\end{thebibliography}

%

\end{document}